\documentclass[12pt,preprint]{aastex}
\usepackage{multirow}

\shorttitle{Discovering the missing 2.2$<$z$<$3 quasars}
\shortauthors{Wu et al.}

\begin{document}

\title{Discovering the missing 2.2$<$z$<$3 quasars by combining optical variability and optical/near-IR colors
\footnote{Observations reported here were obtained at the Bok telescope of Steward Observatory, and at the MMT Observatory, a joint facility of the University of Arizona and the Smithsonian Institution.}}

\author{Xue-Bing Wu\altaffilmark{1,2},
Ran Wang\altaffilmark{3,2},  Kasper B. Schmidt\altaffilmark{4}, Fuyan Bian\altaffilmark{2}, Linhua Jiang\altaffilmark{2}, and Xiaohui Fan\altaffilmark{2}}
\altaffiltext{1}{Department of Astronomy, School of Physics, Peking University, Beijing 100871, China; email: wuxb@bac.pku.edu.cn}
\altaffiltext{2}{Steward Observatory, University of Arizona, Tucson, AZ 85721}
\altaffiltext{3}{Jansky Fellow of the National Radio Astronomy Observatory (NRAO)}
\altaffiltext{4} {Max Planck Institut fuer Astronomie, Koenigstuhl 17, D-69117 Heidelberg, Germany}
\begin{abstract}
The identifications of quasars in the redshift range 2.2$<$z$<$3 are known  
to be very inefficient as their optical colors are indistinguishable from 
those of stars. Recent studies have proposed to use optical variability or near-IR 
colors to improve the identifications of the missing quasars in this redshift 
range. Here we present a case study by combining both factors. We select
a sample of 70 quasar candidates from variables in SDSS Stripe 82, which are 
non-UV excess sources and have UKIDSS near-IR public data. They are clearly separated 
into two parts on the Y-K/g-z color-color diagram, and 59 of them meet or lie
close to a newly proposed Y-K/g-z selection criterion for z$<$4 quasars. 44 of these 59
sources have been previously identified as quasars in SDSS DR7, and 35 among them are
quasars at 2.2$<$z$<$3.  
We present spectroscopic observations of 14 of 15 remaining quasar 
candidates using 
the Bok 2.3m telescope and 
the MMT 6.5m telescope, and  
successfully identify all of them as new quasars at z=2.36 to 2.88. 
 We also apply this method
to a sample of 643 variable quasar candidates with SDSS-UKIDSS nine-band photometric data  
selected from 
1875 new quasar candidates in SDSS Stripe 82 given by Butler \&
Bloom based on the time-series selections, and find that 188 of them
 are probably new quasars with photometric redshifts
at 2.2$<$z$<$3. 
Our results indicate that the combination of optical variability 
and optical/near-IR colors is probably the most efficient way
in finding 2.2$<$z$<$3 quasars and very helpful for constructing a complete quasar sample.
We discuss its
implications to the ongoing and upcoming large 
optical and near-IR sky surveys.
\end{abstract}
\keywords{galaxies: active --- quasars: general 
--- quasars: emission lines }

\section{Introduction}
Since their discovery in 1960s (Schmidt 1963), quasars have become important 
extra-galactic objects in astrophysics. They not only can 
be used to probe the physics of supermassive black holes and accretion/jet 
process, but also are closely related to the studies of galaxy 
evolution, intergalactic medium, large scale structure and cosmology. More than 120,000 quasars have 
been discovered from the large
optical sky surveys, such as the Two-Degree Fields survey (Boyle et al. 2000)
and the Sloan Digital Sky Survey (SDSS, York et al. 2000; Schneider et al. 2010). The
quasar candidates in these surveys were mainly selected by
optical colors, namely that, due to the strong UV and optical emissions,
quasars at $\rm z<2.2$ and $\rm z>3.0$ can be distinguished from
the stellar objects in the color-color and color-magnitude diagrams based on optical
photometry (Smith et al. 2005; Richards et al. 2002; Fan et al. 2000).
However, in the redshift range $\rm 2.2<z<3.0$, the redshifted
spectral energy distributions of quasars show similar optical
colors to that of normal stars, and quasar selections using the optical 
color-color diagrams 
become very inefficient due to the serious contaminations
of stars (Fan 1999; Richards et al. 20002; 2006; Scheneider et al. 2007). Because of the crucial 
importance of
z$>$2.2 quasars in studying the Ly$\alpha$ forest and cosmic baryon acoustic 
oscillation
(BAO) (White 2003; McDonald \& Eisenstein 2007) and in constructing the accurate 
luminosity function to study the 
quasar evolution in the mid-redshift universe (Wolf et al. 2003; Jiang et al. 2006),  
we have to explore other efficient ways to 
identify the missing $\rm 2.2<z<3.0$ quasars.

In the last a few years, two main approaches have been taken to separate quasars
and stars rather than using optical color-color diagrams. The first approach is
to use optical variability, as this is one of the well known quasar properties
(Hook et al. 1994; Cristiani et al. 1996; Giveon et al. 1999).
Schmidt et al. (2010) have proposed a method to select quasar candidates by 
their intrinsic
variability. They showed that the quasar structure functions, constructed from 
the light-curves of known quasars in SDSS Stripe 82 (hearafter S82; see also Sesar et al. 2007),
can be modeled by a power-law function with amplitude A and power-law
index $\gamma$. Quasars can be separated from  stars in
the $\rm A-\gamma$ plane, which enables efficient selection
of quasar candidates based on long-term
single-band optical photometry (Schmidt et al. 2010). They also pointed out
that in the redshift range $\rm 2.5<z<3.0$, variability can  help to select 
quasars with a completeness of 90$\%$. 
MacLeod et al. (2011) also developed a method to use the damping timescale and
asymptotic amplitude of variable sources in S82 to separate quasars from stars
with an efficiency higher than 75\%.
Butler \& Bloom (2011) recently presented a similar
time-series study of quasars in S82, and proposed to use
two statistics, a quasar-like variability metric and a non-quasar variability
metric, to yield the separation of quasar candidates from stars. They claimed 
that with their method they can achieve nearly a factor of two increase of quasars at
 $\rm 2.5<z<3.0$. In addition,  very recent results from the SDSS-III Baryon Oscillation 
Spectroscopic Survey (BOSS; Eisenstein et al. 
2011) also
confirmed the high success rate of spectroscopically identifying variability 
selected quasars, which leads to a significant increase 
of z$>$2.2 quasar density  in S82 than that based on optical colors only
(Palanque-Delabrouille et al. 2011; Ross et al. 2011).

The second approach to separate z$>$2.2 quasars from stars is to utilize
their near-IR colors. As the continuum emission from
stars usually decreases more rapidly from optical to the near-IR wavelengths than
that of quasars, the near-IR colors for stars are different from quasars. This
leads to a method to use the K-band excess to identify quasars at z$>$2.2 
(eg. Warren, Hewett \& Foltz 2000; Croom, Warren \& Glazebrook 2001; Sharp et al. 2002; 
Hewett et al. 2006; Chiu et al. 2007;
Maddox et al. 2008;  Smail et al. 2008; Wu \& Jia 2010).  Using the photometric data in the
$ugriz$ bands of SDSS DR7 (Abazajian et al. 2009) and YJHK bands of UKIRT InfraRed Deep Sky Surveys (UKIDSS\footnote{
The UKIDSS project is defined in Lawrence et al. (2007). UKIDSS uses the UKIRT Wide Field Camera (WFCAM; Casali et al. 2007) and a photometric system described in Hewett et al. (2006). The pipeline processing and science archive are described in Hambly et al. (2008).}) Large Area Survey (LAS)
 DR3, Wu \& Jia (2010) compiled a
sample of 8498 SDSS-UKIDSS quasars and a sample of 8996 SDSS-UKIDSS stars. Based on
these two samples they compared different optical/near-IR color-color diagrams
and proposed an efficient empirical criterion for selecting  z$<$4
 quasars 
in the near-IR Y$-$K and optical g$-$z color-color
diagram (i.e. $\rm Y-K>0.46(g-z)+0.53$, where all magnitudes are Vega 
magnitudes). With this criterion, they obtained the
completeness of 98.6$\%$ of recovering z$<$4 quasars with the mis-identifying
rate of 2.3$\%$ of classifying stars as quasars. A check with the FIRST 
(Becker, White \& Helfand 1995) radio-detected 
SDSS quasars, which are believed to be free of color selection bias, 
also proved that with this Y-K/g-z criterion they can achieve the 
completeness higher than 95$\%$ for these radio-detected quasars with z$<$3.5, which seems 
to be difficult in the case of using the SDSS optical color selection criteria 
alone where a dip around z$\sim$2.7 in the redshift distribution obviously exists
(Richards et al. 2002,2006; Scheneider et al. 2007,2010). 
Recently, Peth, Ross \& Schneider (2011) extended the
study of Wu \& Jia (2010) to a much larger sample of 130000 SDSS-UKIDSS selected
quasar candidates and re-examined the methods of separating stars and 
mid-redshift quasars with the near-IR/optical colors. 
Using the Y-K/g-z selection criterion, Wu et al. 
(2010a,b) also successfully identified some $\rm 2.2<z<3.0$ quasars during the 
commissioning
period of the Chinese GuoShouJing Telescope (LAMOST), which provides further
supports to the effectiveness of selecting the mid-redshift quasars using the 
optical/near-IR colors.

Although both approaches we mentioned above can be used to identify quasars
at  $\rm 2.2<z<3.0$, a more ideal approach is to combine the variability and
optical/near-IR color to achieve the
maximum efficiency. In this paper,  we present a case study by selecting
a sample of variable, non-UV excess, SDSS-UKIDSS quasar candidates in S82 (Schmidt et al. 2010), and  
spectroscopically identifying 14 new quasars 
at z=2.36 to 2.88. 
 We also apply this method
to some new variable quasar candidates in S82 recently 
suggested by Butler \&
Bloom (2011) and found that 188 SDSS-UKIDSS sources are probably new quasars with 2.2$<$z$<$3. 
We describe the sample 
selections and spectroscopic observations in Section 2, present more new  2.2$<$z$<$3 quasar candidates in S82 in Section 3 and discuss the results 
in Section 4.

\section{Target selections and spectroscopic observations}
Our purpose is to efficiently select $\rm 2.2<z<3.0$ quasars by combining
the variability and optical/near-IR colors, so we focus on S82 region where both
variability and SDSS-UKIDSS photometric data are available with high quality. 
A sample of 118 non-UV excess quasar candidates from S82 
has been selected with the algorithm presented in Schmidt et al. (2010), which have
UV-optical colors similar to that of stars (i.e. consistent
with the observed colors of quasars at $\rm 2.2<z<3.0$ and optical
variability properties consistent with the region defined by quasars
on the $\rm A-\gamma$ plane).
70 of them have near-IR YJHK photometric data from
the UKIDSS/LAS DR4\footnote{Available at http://surveys.roe.ac.uk/wsa/}. 
All photometric magnitudes are corrected for Galactic extinction using a map of
Schlegel, Finkbeiner \& Davis (1998).
We plot the 70 objects on the Y-K/g-z color-color diagram (see Fig. 1). 
They are clearly separated into two parts on this diagram. 54 sources
match the selection criterion of $\rm Y-K>0.46(g-z)+0.82$ defined for z$<$4 quasars (here we convert the 
original criterion given in Wu \& Jia (2010) to a new one to keep the 
g and z magnitudes in AB system and Y and K magnitudes in Vega system, 
see dashed line in
Fig. 1). Five sources locate slightly below  but very close to the  criterion.
Therefore, we think these 59 sources are probably $\rm 2.2<z<3.0$ quasars.
The photometric redshifts of these 59 sources are estimated to 
be from z=2.43 to 3.05 using their nine-band SDSS-UKIDSS photometric data
with a program introduced in Wu \& Jia (2010). 
Indeed, 44 among them have been spectroscopically identified as quasars by SDSS previously. 
These 44 known quasars have spectroscopic redshifts from 0.59 to 3.29, and
35 of them are $\rm 2.2<z<3.0$ quasars. The
spectroscopic redshifts for 40 of these 44 known quasars are consistent with
their photometric redshifts within $\rm |\Delta z|\leq 0.3$. This confirms
the high efficiency of selecting $\rm 2.2<z<3.0$ quasars by combining the
variability and optical (g-z)/near-IR (Y-K) color. Spectroscopic
identification of the remaining 15 quasar candidates is needed.

Apart from the above 59 quasar candidates, the other 11 objects are located much 
below the quasar selection criterion in Fig. 1, and  their Y-K and g-z colors are 
indistinguishable from those of stars in the stellar locus (see Fig. 5 of Wu \& Jia (2010)). In addition,
ten of them have very bright optical magnitudes (i.e. $i<16.5$) and are unlikely to be 
quasars at the expected redshifts ($\rm 2.2<z<3.0$). Indeed, four of them 
(SDSS J034751.14-001730.7, SDSS J035208.92+005919.6, SDSS J224630.25+010018.3 and
SDSS J225342.13+011207.1) have already 
been cataloged as stars in the SIMBAD database\footnote{http://simbad.u-strasbg.fr/simbad/}.  

As we mentioned above, spectroscopic identification is still required for the remaining 15 
 quasar candidates with $\rm 2.2<z<3.0$ in S82. All of them have $i$-band magnitudes brighter
than 19.3. 
In this paper we present optical spectra for 14 of them\footnote{The only one left is
SDSS J220808.97+002858.3 with a photometric redshift of 2.78. We are planning to observe it
in the fall of 2011.}. 
Eight of them were observed using the Boller \& Chivens
Spectrograph on the Bok 2.3m Telescope at Kitt Peak in November 2010. The observation 
covers a wavelength range of 3620--6900 $\rm \AA$ with a spectral 
resolution of 8.3 $\rm \AA$. The spectra of the other six objects were 
obtained with the Blue Channel Spectrograph on the MMT 6.5m Telescope at Mt. Hopkins
in December 2010, 
with a wavelength coverage of 3600 $\rm \AA$ to 8000 $\rm \AA$ and a spectral 
resolution of 5.8 $\rm \AA$. We reduce the data with IRAF package and some 
broad line emissions, 
such 
as Ly$\alpha$+$\rm N\,{\small V}$, $\rm Si\,{\small IV}$+$\rm O\,{\small IV}]$, 
and $\rm C\,{\small IV}$, have been clearly detected in the spectra 
of all of 14 quasar candidates. 
We measure the redshifts of these 14 new quasars by 
fitting Gaussian line profiles to the Ly$\alpha$+$\rm N\,{\small V}$,  
$\rm Si\,{\small IV}$+$\rm O\,{\small IV}]$ $\rm \lambda1399$ 
and $\rm C\,{\small IV}$ $\rm lambda1549$ emission lines.
The details of the sources and observational results, including their
names, coordinates, magnitudes, exposure times, photometric and spectroscopic
redshifts,  are summarized in Table 1. The spectra 
of these 14 new quasars with z=2.36 to 2.88 taken with Bok and MMT are presented in Fig. 2 and 
Fig. 3, respectively. 
These observations clearly demonstrated the high efficiency of selecting 
$\rm 2.2<z<3.0$ quasars by combining the variability and optical/near-IR colors.

For the 11 sources located much below the quasar selection
criterion of the Y-K/g-z color-color diagram, we also took the spectra of one 
of them (SDSS J035658.21+003801.8, $i$=18.69) with the Bok 2.3m telescope
in November 2010 and two of them (SDSS J034950.99+ 010845.9, $i$=11.63 and 
SDSS J035816.05+002351.9, $i$=13.58) 
with the 2.16m telescope of Xinglong/NAOC in January 2011,
and confirmed their nature as stars due to the lack of emission lines
and the presence of Balmer absorption features. Together with other four previously known 
stars, seven of these 11 sources located in the stellar locus have been identified 
as stars. Although there are still four sources 
remaining unidentified, they are obviously too bright ($i<14$) to be quasars. Therefore, 
we believe that all these 11 sources are stars with certain level of optical variability.
Combining with their optical/near-IR colors we can easily separate them from quasars.

\section{More $\rm 2.2<z<3.0$ quasar candidates in SDSS Stripe 82}
In a recent paper, Butler \& Bloom (2011) presented a similar
time-series study of quasars in S82 as in Schmidt et al. (2010) and MacLeod et al. (2011). 
They proposed
two different statistics, namely a quasar-like variability metric and a non-quasar variability
metric, to separate quasar candidates from stars. They obtained 1875 new quasar candidates in
S82 and claimed 
that with their method they can achieve nearly a factor of two increase of quasars at
 $\rm 2.5<z<3.0$. Here we use their variable quasar candidates to cross-correlate with
the sources in the UKIDSS/LAS DR5 and obtained 643 new quasar candidates with SDSS-UKIDSS 
nine-band photometric data. In Fig. 4 we plot these sources in the Y-K/g-z diagram, in
comparison with the quasar selection criterion suggested by Wu \& Jia (2010). 597 of these
643 sources (with a fraction of 93$\%$) meet the selection criterion, suggesting that most of them should
be real quasars with z$<$4. This comparison also provides mutual support to the quasar
selection method based on variability or optical/near-IR colors. 

For more reliably selecting $\rm 2.2<z<3.0$ quasars from these 597 quasar candidates, we used a program 
introduced in Wu \& Jia (2010) to estimate the photometric redshifts of quasar candidates based
on their SDSS-UKIDSS nine-band photometric data. In the upper panel of Fig. 5 we show the photometric
redshift distribution of these 596 new quasar candidates. Although they distribute in a
broad redshift range from 0.1 to 3.8, obviously a large fraction of them are at $\rm 2.2<z<3.0$. 
Among these  597 quasar candidates, 244 sources have photometric redshifts larger than 2 and
188 of them are $\rm 2.2<z<3.0$ quasar candidates. Considering the fact that only 948  
 quasars at $\rm 2.2<z<3.0$ in S82 have been identifed in the SDSS DR7 (Schneider et al. 2010), the
fraction of $\rm 2.2<z<3.0$ quasar candidates in our SDSS-UKIDSS variable source sample is
significantly higher. This is understandable because SDSS quasar survey mainly focused on
finding quasars with z$<$2.2 and z$>$3.5 (Richards et al. 2002). Many quasars with 
$\rm 2.2<z<3.0$ are therefore missing in the SDSS quasar survey but can be discovered
by combining variability and optical/near-IR colors as we suggested in this paper.

In the lower panel of Fig. 5 we show the distribution of the dereded $i$-band magnitudes
of 597 quasar candidates, as well as that of 188 quasar candidates at $\rm 2.2<z<3.0$. Clearly,
majority of them are located between $i=19.1$ and $i=20.5$. This is also the reason why
they are missing in the SDSS survey because most of the SDSS known quasars have $i<19.1$. 
We expect that the ongoing BOSS survey in SDSS III,
which aims to discover 150000 quasars with z$>$2.2 (Eisenstein et al. 
2011; Ross et al. 2011), could confirm the quasar nature
and redshifts of these quasar candidates soon. In Table 2 we listed the coordinates, 
photometric redshifts, and  SDSS and UKIDSS magnitudes  
of the 188  quasar candidates at $\rm 2.2<z<3.0$. We also noticed that three bright sources ($i<19.3$)
among
 them have been
spectroscopically identified by us in Section 2.

\section{Discussion}
We have presented a case study to demonstrate that we can effectively select  $\rm 2.2<z<3.0$ quasars by combining the optical 
variability and optical/near-IR colors. Our successful spectroscopic identifications of 14 new quasars at
  z=2.36 to 2.88 with the Bok 2.3m telescope and the MMT 6.5m telescope 
provide further support to this combination approach, which can be used to select
quasars with probably the highest efficiency (here we define the efficiency as the percentage of
quasars identified from the spectroscopic targets, similar to the definition in SDSS-III (Ross et al. 2011)).
We also compiled a catalog of 188 quasar candidates with photometric redshifts at $\rm 2.2<z<3.0$ 
 from variable SDSS-UKIDSS
sources in S82, and expect that the ongoing SDSS III spectroscopy will confirm their quasar nature 
and redshifts soon.

We noticed that although combining the optical/near-IR colors and time-series information can help 
increase the efficiency in identifying quasars, it may decrease the completeness of quasars if both
selection criteria on colors and variability are required. This can also be seen from Fig. 1 and 
Fig. 4 as some quasars selected by variability do not meet our color selection criterion.
One possible way to avoid this is to decrease the threshold for each criterion. For example,
relaxing our color criterion to $Y-K>0.46(g-z)+0.50$ would include all variability-selected quasars
in Fig. 1 and 98.8\% of variability-selected quasar candidates in Fig. 4, without increasing much
contamination from stars as most of them are less variable and still far away from our color 
selection criterion. However, how to best combine both the optical/near-IR colors and time-series information to select quasars with both higher efficiency and higher completeness obviously needs more investigations in the future based on complete samples of quasars and stars in
certain sky areas.  

For using this combination approach, we need both the optical variability measurements 
and optical/near-IR photometry in a large sky area. 
Especially for finding z$>2.2$ quasars, deeper imaging and multi-epoch photometry are neccesary.  
However, so far both variability and optical/near-IR photometric observations have been realized only for
a small part of the sky, such as in S82. Because the typical variability timescales of quasars are
usually in years in the optical band, we need to measure the variability of sources for many epochs in at least several years 
in order to get better statistics to determine their variability features. That is why so far the variability
studies related to quasars have been done only on some smaller sky areas, which significantly 
limits the efforts in discovering quasars with variability. 
However, even if we may not have both time-series and color information for quasar candidate selection in most sky areas,  utilizing both information as much as possible often allows us to get the most quasars.
Fortunately, there are also several ongoing 
and upcoming large projects with both photometric and variability information, especially the Panoramic Survey Telescope \& Rapid Response System (Pan-STARRS;
Kaiser et al. 2002) and the Large Synoptic Survey Telescope (LSST; Ivezic et al. 2008).
The multi-epoch photometry in multi-bands covering a large part of the sky by these facilities
will hopefully provide better opportunities to use variability to construct a much larger sample of 
quasars than that currently available.

On the other hand, several ongoing and upcoming optical and near-IR photometric sky surveys will also provide
crucial helps to us to extend the SDSS-UKIDSS optical/near-IR color selection of quasars to larger
and deeper fields. In addition to SDSS III (Eisenstein et al. 2011), which has taken 2500 deg$^2$ 
further imaging
in the south galactic cap, the SkyMapper (Keller et al. 2007) and Dark Energy Survey (DES; The Dark Energy Survey Collaboration 2005) will also present the multi-band optical 
photometry in 20000/5000 deg$^2$ of the southern sky, with the magnitude limit of 22/24 mag in $i$-band,
respectively. The Visible and Infrared Survey Telescope for Astronomy (VISTA; Arnaboldi et al. 2007) will 
carry out its
VISTA Hemisphere Survey (VHS) in the near-IR YJHK bands for 20000 deg$^2$ of the southern sky with
a magnitude limit at K=20.0, which is about five magnitude and two magnitude deeper than the Two Micron ALL
Sky Survey (2MASS; Skrutskie et al. 2006)
and UKIDSS/LAS limits (Lawrence et al. 2007), respectively. Therefore, the optical and near-IR photometric data
obtained with these ongoing and upcoming surveys will provide us a large database for quasar selections. 
By combining the optical variability and the optical/near-IR colors, we expect that a much larger and more
complete quasar sample can be efficiently constructed in the near future.

Although by combining the variability and optical/near-IR colors we can efficiently select
quasar candidates and reliably estimate their photometric redshifts, the spectroscopic 
identifications are still crucial to determine their quasar nature and redshifts.
The ongoing BOSS project in SDSS III has identified 29000 quasars with $z>2.2$ and expects to obtain
the spectra of 150000 quasars at  $2.2<z<4$ (Eisenstein et al. 2011; Ross et al. 2011). We believe that many      
  $\rm 2.2<z<3.0$ quasars, including the candidates we listed in this paper,  should be 
spectroscopically identified by BOSS. In addition, 
the Chinese GuoShouJing telescope (LAMOST; Su et al. 1998), a spectroscopic telescope with 4000 fibers
currently in the commissioning phase, 
is also aiming at discovering 0.3 million quasars
with magnitudes bright than $i=20.5$ (Wu et al. 2010b). 
By combining the variability and optical/near-IR colors, large input catalogs of reliable quasar candidates
will be provided to these quasar surveys for future spectroscopic observations.
Therefore, we expect that a much larger and more complete quasar sample covering a wider range of
 redshift  will be constructed in the near future, which will play an important role
in studying extra-galactic astrophysics, including  the physics of accretion around supermassive
black holes, galaxy  
evolution, intergalactic medium, large scale structure and cosmology.

\acknowledgments 
We thank Zhaoyu Chen and Wenwen Zuo for taking the spectra of two star 
candidates with the 2.16m telescope at Xinglong/NAOC. 
XBW is supported by the National Natural Science Foundation of China (11033001) and the National Key Basic Research Science Foundation of China (2007CB815405). He thanks the colleagues at Steward Observatory, University of Arizona
for hospitality during his visit there in the spring of 2011 as a senior visiting fellow supported by the
Chinese Scholarship Council. RW acknowledges the support from the National Radio Astronomy Observatory (NRAO) through the Jansky Fellowship program.
NRAO is a facility of the National Science Foundation operated under cooperative agreement by Associated Universities, Inc. KBS is funded by and would like to thank the Marie Curie Initial Training Network ELIXIR, which is
funded by the Seventh Framework Programme (FP7) of the European Commission. KBS is a member of the International Max Planck Research School for Astronomy and Cosmic Physics at the University of Heidelberg (IMPRS-HD), Germany. 
This work was partially supported by the Open Project Program of the Key Laboratory of Optical Astronomy, NAOC, CAS.
This research has made use of the SIMBAD database,
operated at CDS, Strasbourg, France. 

Funding for the SDSS and SDSS-II has been provided by the Alfred P. Sloan Foundation, the Participating Institutions, the National Science Foundation, the US Department of Energy, the National Aeronautics and Space Administration, the Japanese Monbukagakusho, the Max Planck Society and the Higher Education Funding Council for England. The SDSS web site is http://www.sdss.org/.

The SDSS is managed by the Astrophysical Research Consortium for the Participating Institutions. The Participating Institutions are the American Museum of Natural History, Astrophysical Institute Potsdam, University of Basel, University of Cambridge, Case Western Reserve University, University of Chicago, Drexel University, Fermilab, the Institute for Advanced Study, the Japan Participation Group, Johns Hopkins University, the Joint Institute for Nuclear Astrophysics, the Kavli Institute for Particle Astrophysics and Cosmology, the Korean Scientist Group, the Chinese Academy of Sciences (LAMOST), Los Alamos National Laboratory, the Max-Planck-Institute for Astronomy (MPIA), the Max-Planck-Institute for Astrophysics (MPA), New Mexico State University, Ohio State University, University of Pittsburgh, University of Portsmouth, Princeton University, the United States Naval Observatory and the University of Washington.

{\it Facilities:} \facility{Sloan (SDSS)}, \facility{UKIDSS}, \facility{Bok}, \facility{MMT},\facility{2.16m/NAOC}

\begin{table}
{\scriptsize \caption{Parameters of 14 new $\rm 2.2<z<3.0$ quasars in S82}
\begin{tabular}{crrccccc}
\hline \noalign{\smallskip}
Name & RA & Dec & i &  $\rm z_{photo}$ & $\rm z_{spec}$ & Exposure Time & Telescope \\
(SDSS) & degree& degree  &    &   &  & s &  \\
\noalign{\smallskip} \hline \noalign{\smallskip}
J000050.59+010959.1 & 0.21081 & 1.16644    & 19.22 & 2.58 &2.37  &3600& BoK \\ 
J000121.87$-$000327.1 & 0.34113 & $-$0.05754 & 18.47 & 2.78 &2.88 &300&MMT \\
J002117.11$-$002841.7 & 5.32131 & $-$0.47824 & 18.68 & 2.68 &2.85 &300&MMT \\
J013450.27+003537.1 & 23.70948 & 0.59367 & 17.52 & 2.68 &2.69  & 3600 & Bok \\
J022836.08+000939.2 & 37.15035 & 0.16091 & 18.34 & 2.68 &2.63  & 3600& Bok \\
J034025.90+000807.6 & 55.10792 & 0.13545 & 19.08 & 2.63 &2.64   &3600& Bok \\
J034008.54+010714.8 & 55.03557 & 1.12081 & 18.88 & 2.83 &2.84   & 2400 & Bok\\
J034337.67$-$000350.2 & 55.90698 & $-$0.06395&18.14&2.88&2.85 &3600& Bok\\
J214633.34+000318.5 & 326.63895 &0.05516 &17.83 & 2.83 &2.79  &600 & MMT\\
J221602.32+005826.5 & 334.00967 &0.97406 &17.86 & 2.83 &2.83  &1200&Bok\\
J225257.56+004524.0 &343.23984  &0.75669 &19.25 & 2.83 &2.74   &900&MMT \\
J225355.31+005146.1 &343.48050  &0.86281 &19.24 & 2.43 &2.37   &1800&Bok\\
J231302.58+004105.1 &348.26074  &0.68475 &19.17 & 2.73 &2.63   &600 &MMT\\
J233659.54+003843.5 &354.24808  &0.64543 &18.89 & 2.68 &2.72   &300 &MMT\\
\noalign{\smallskip} \hline
\end{tabular}\\
}
\end{table}

\begin{figure}
\plotone{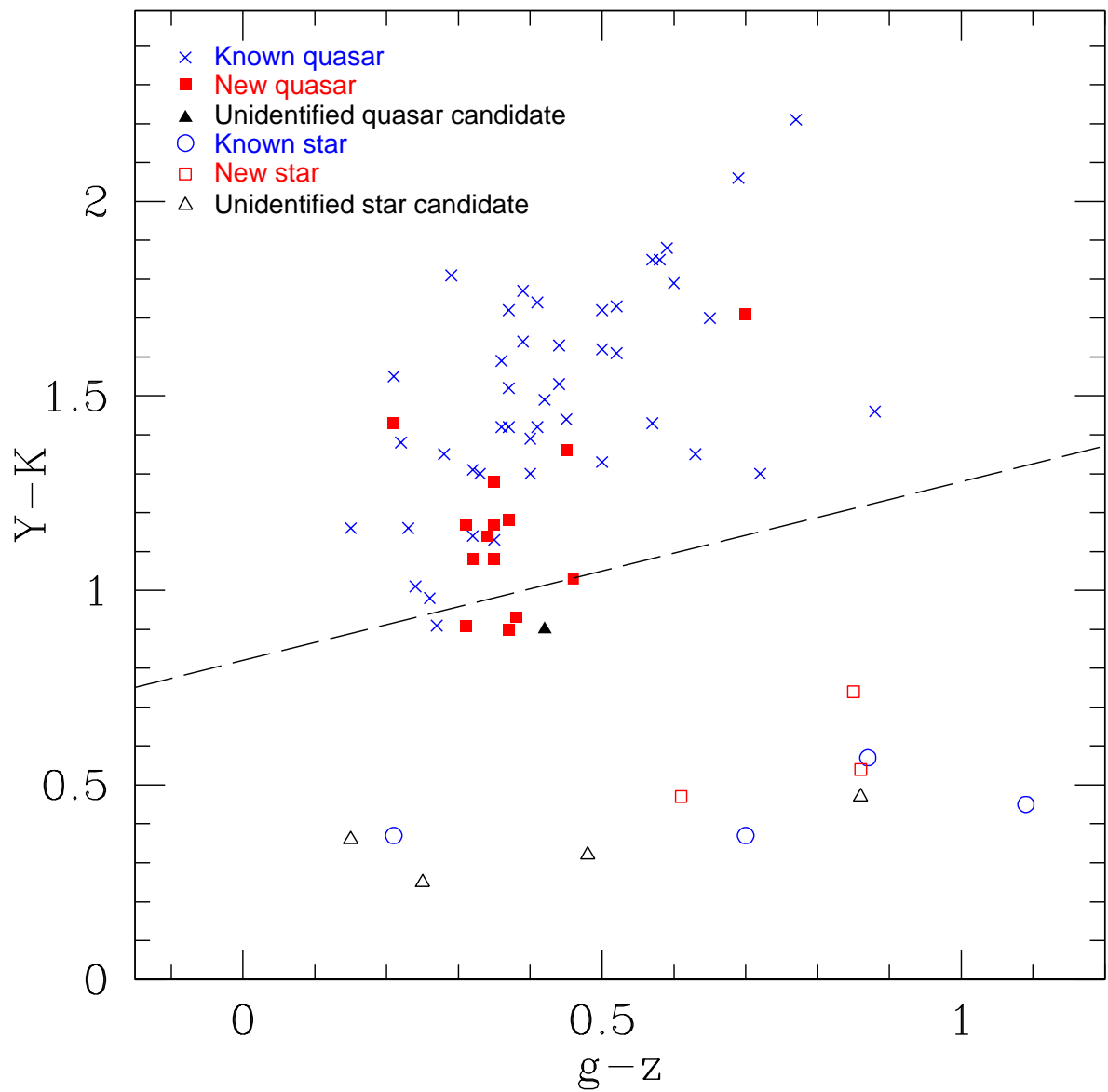}
\caption{The Y-K and g-z colors of 70 SDSS-UKIDSS non-UV excess selected variable quasar candidates 
from Schmidt et al. (2010), in comparison with the z$<$4 quasar selection criterion (dashed line), Y-K$>$0.46*(g-z)+
0.82 (where g and z in AB magnitudes and Y and K in Vega magnitudes, proposed by Wu \& Jia (2010). Filled
squares represent 14 new 2.2$<$z$<$3 quasars identified in this paper. Crosses and a filled triangle 
represent the previously known SDSS quasars and a probable quasar without spectroscopy, respectively.
Open circles, squares and triangles denote the known stars, stars identified in this paper and probable
stars without spectroscopy, respectively.}
\end{figure}

\begin{figure}
\plotone{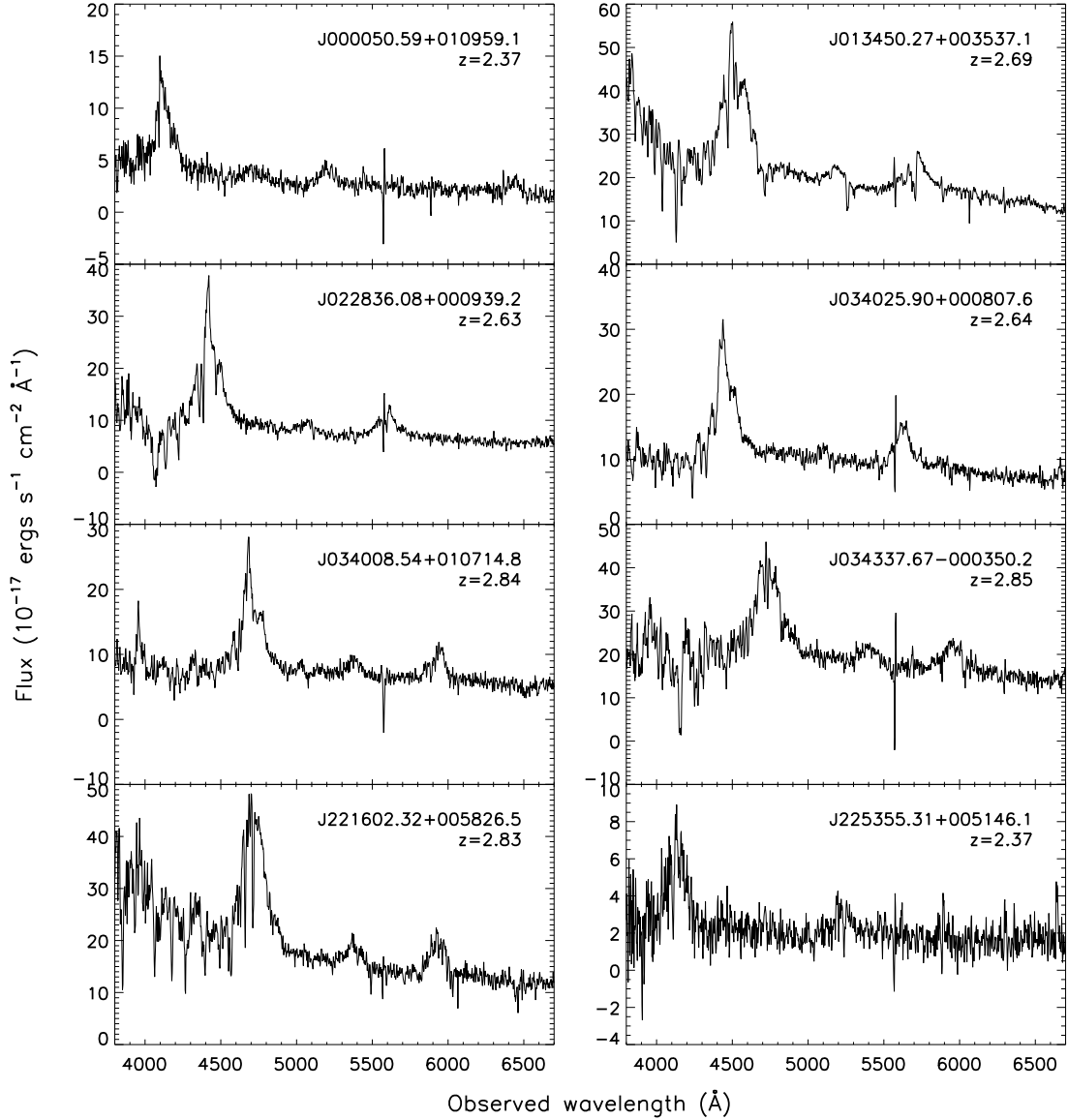}
\caption{Bok spectra of the eight new quasars at $2.2<z<3$ selected by
combination of variability and optical/near-IR color. The strongest emission line
in each spectrum is Ly$\alpha$+$\rm N\,{\small V}$.}
\end{figure}

\begin{figure}
\plotone{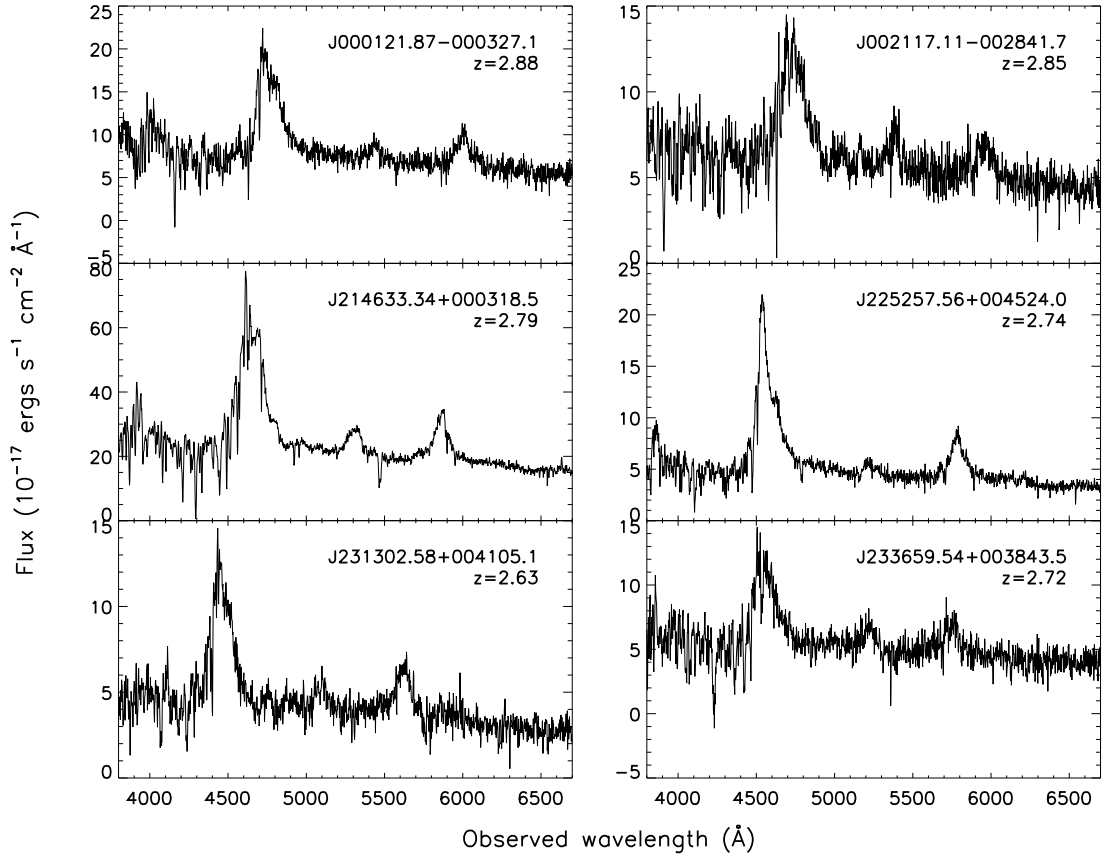}
\caption{MMT spectra of the six new quasars at $2.2<z<3$ selected by
combination of variability and optical/near-IR color.}
\end{figure}

\begin{figure}
\plotone{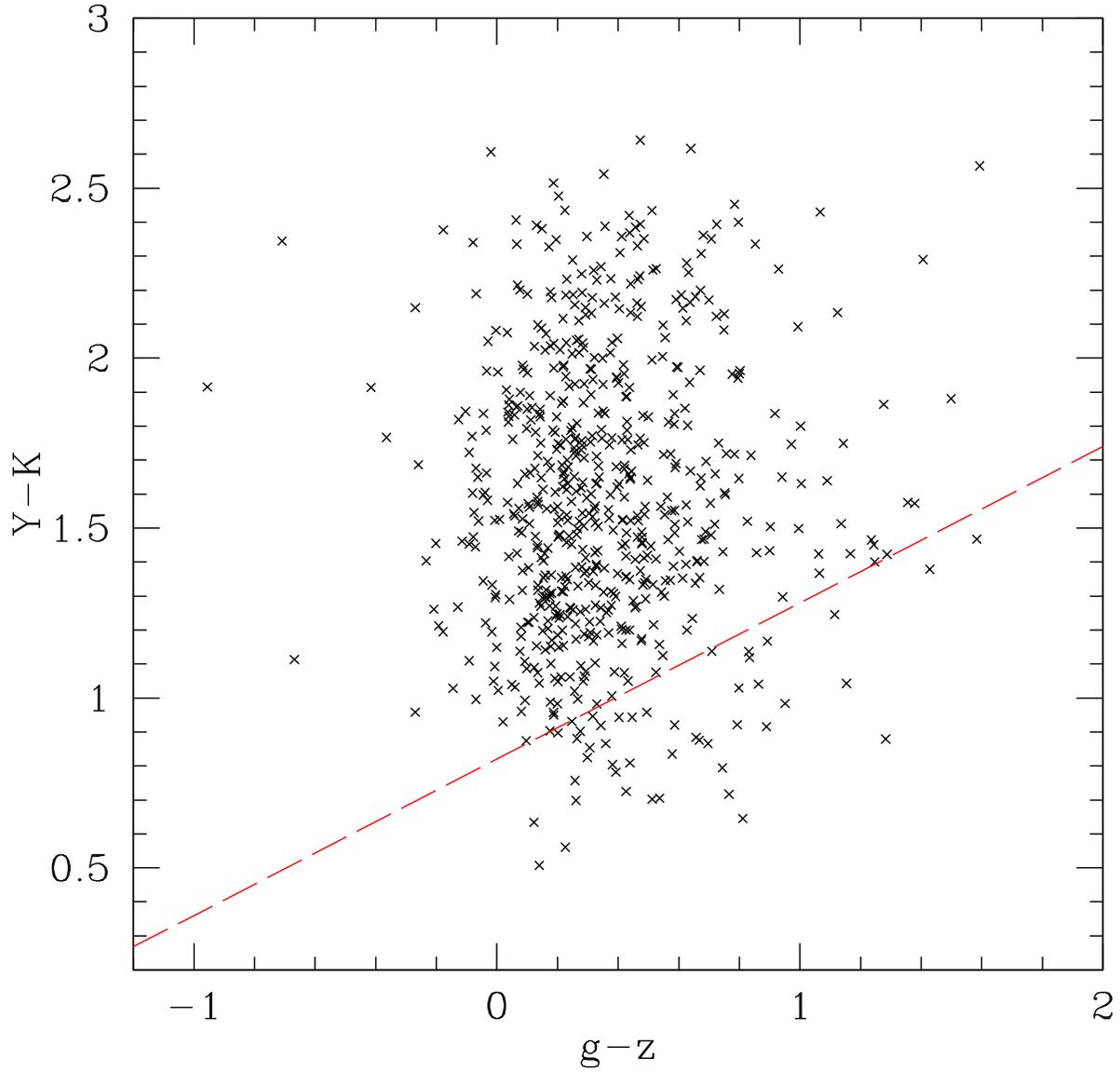}
\caption{The Y-K and g-z colors of 643 variable quasar candidates with SDSS-UKIDSS nine-band
photometric data, selected from 1875 new quasar candidates in Butler \& Bloom (2011). 
597 of them meet  the 
z$<$4 quasar selection criterion (shown as dashed
line) given by Wu \& Jia (2010).}
\end{figure}

\begin{figure}
\plotone{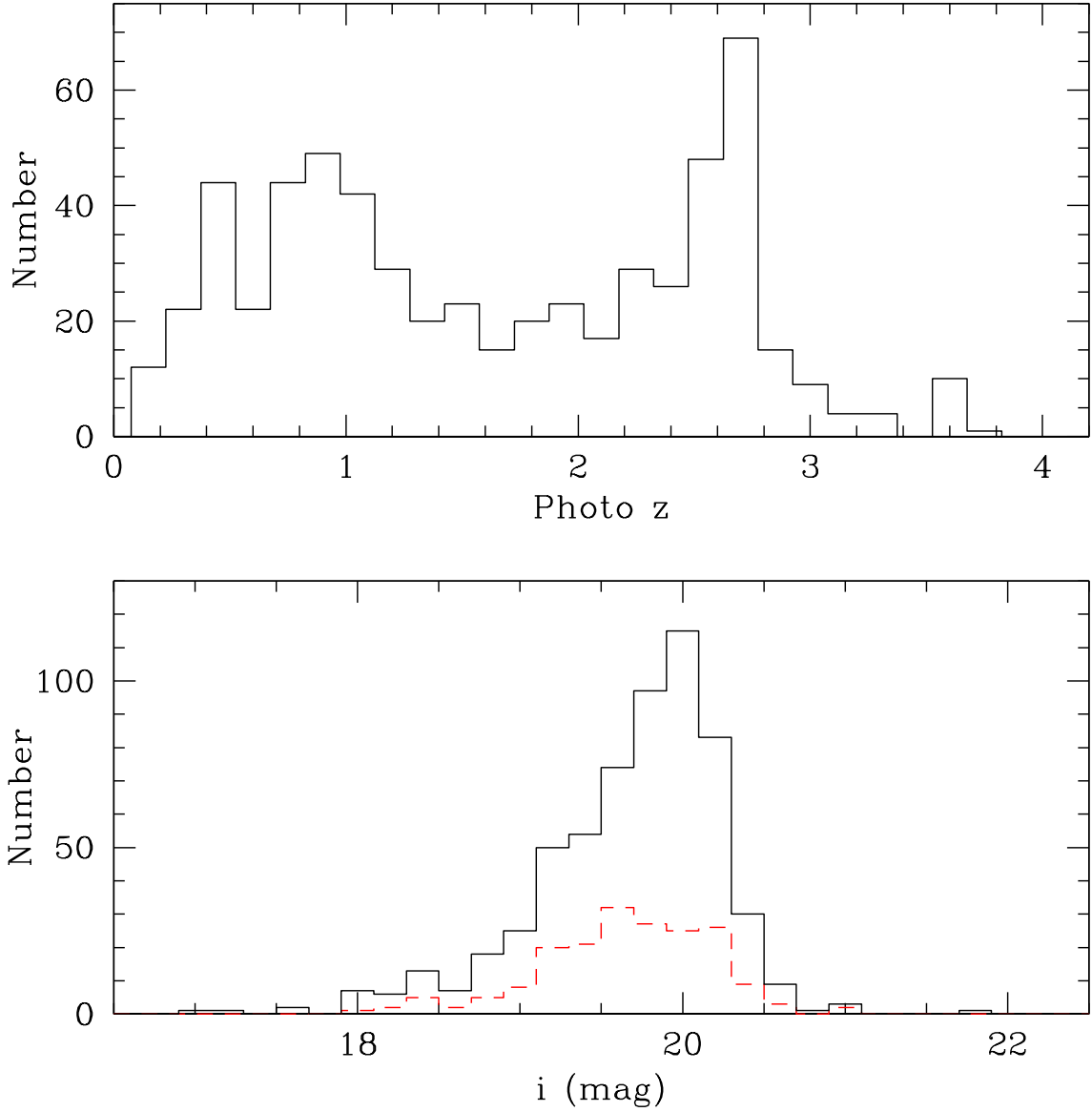}
\caption{Upper panel: The distribution of photometric redshifts of 597 variable SDSS-UKIDSS quasar candidates
satisfying the Y-K/g-z selection criterion of z$<$4 quasars. A large fraction of sources are at 2.2$<$z$<$3. 
Lower panel: The distribution of SDSS dereded $i$ magnitudes of 597 variable SDSS-UKIDSS quasar candidates.
The dashed histogram represents the distribution of 188 quasar candidates with photometric redshifts of 2.2$<$z$<$3. }
\end{figure}

\begin{table}
{\scriptsize \caption{188 quasar candidates with $\rm 2.2<z<3.0$ in S82}
\begin{tabular}{cccccccccccc}
\noalign{\smallskip}
\hline \noalign{\smallskip}
 RA & Dec & $\rm z_{photo}$ & u&g&r&i&z&Y&J&H&K  \\
 degree& degree  &    &   &  & &   &  &\\
\noalign{\smallskip} \hline \noalign{\smallskip}
0.21081127 & 1.16643864 &    2.575& 19.731& 19.138&  19.090& 19.181& 18.985& 18.294& 18.142& 17.669& 16.891\\
0.34113037 &$-$0.05754064 &   2.825& 19.751& 18.737&  18.499& 18.386& 18.452& 17.578 &17.212& 16.799& 16.529\\
0.42223178& $-$0.29444873 &   2.425& 20.592& 19.933&  19.786& 19.754& 19.469& 18.910& 18.494& 18.290& 17.374\\
0.44474748& 0.66631052  &  2.775 &20.408& 19.471 & 19.249& 19.215& 19.244 &18.437& 18.166& 17.645 &17.079\\
0.50900025& $-$0.70452072 &   2.425& 21.735& 20.360&  20.146& 20.089& 19.735& 19.245& 18.777& 18.296& 17.136\\
0.54000096& $-$0.03205093 &   2.675& 20.864& 19.865&  19.728& 19.802& 19.773& 18.913& 18.825& 18.170& 17.920\\
0.54369966& $-$1.02404081 &   2.775& 21.279& 19.679&  19.220& 19.206& 19.267& 18.204& 17.953& 17.524& 17.044\\
0.55005872& 0.54818141 &   2.675 &20.648& 19.664&  19.610 &19.555& 19.472& 18.729 &18.538& 18.192 &17.671\\
0.90385887&$-$1.17817945&    2.325& 20.304& 19.843&  19.953& 19.934& 19.889& 19.346& 19.307& 18.910& 18.002\\
1.06989716& 0.24487200  &  2.475 &21.075& 20.298 & 20.305 &20.353 &20.079 &19.190 &19.119 &18.678& 17.877\\
\noalign{\smallskip} \hline
\noalign{\smallskip}
\end{tabular}\\
}
{\scriptsize Note: The SDSS ugriz magnitudes are AB magnitudes and the UKIDSS YJHK magnitudes are Vega magnitudes.
Only a portion of the table is shown here. The whole table will be available online.}
\end{table}


\begin{thebibliography}{}
\bibitem[Abazajian et al. (2009)]{a09}Abazajian, K., et al., 2009, \apjs, 182, 543 
\bibitem[Arnaboldi et al. (2007)]{a07}Arnaboldi, M., et al., 2007, The Messenger, 127,  28
\bibitem[Becker, White \& Helfand (1995)]{b95}Becker, R.H., White, R.L., \& Helfand, D.J., 1995, \apj, 450, 559
\bibitem[Boyle et al. (2000)]{boyle00}Boyle, B.J., et al., 2000, MNRAS, 317, 1014
\bibitem[Butler \& Bloom (2011)]{bb11}Butler, N.R., \& Bloom, J.S., 2011, \aj, 141,93
\bibitem[Casali et al. (2007)]{c07}Casali, M., et al., 2007, A\&A, 467, 777
\bibitem[Chiu et al. (2007)]{cr07}Chiu, K., Richards, G.T., Hewett, P.C., Maddox, N., 2007, MNRAS, 375,1180
\bibitem[Cristiani et al. (1996)]{c96}Cristiani, S., et al., 1996, A\&A, 306, 395
\bibitem[Croom, Warren \& Glazebrook (2001)]{c01}Croom, S.M., Warren, S.J., \& Glazebrook, K., 2001, \mnras, 328, 150
\bibitem[Eisenstein et al. (2011)]{e11}Eisenstein, D., et al. 2011, \apj ~submitted, arXiv:1101.1529
\bibitem[Fan (1999)]{f99}Fan, X., 1999, \aj, 117, 2528
\bibitem[Fan et al. (2000)]{f00}Fan, X., et al. 2000, AJ, 120, 1167
\bibitem[Giveon et al. (1999)]{g99}Giveon, U., et al. 1999, \mnras, 306, 637
\bibitem[Hambly et al. (2008)]{h08}Hambly, N., et al., 2008, MNRAS, 384, 637
\bibitem[Hewett et al. (2006)]{h06}Hewett, P.C., Warren, S.J., Leggett S.K., Hodgkin S.T. 2006, MNRAS, 367, 454
\bibitem[Hook et al. (1994)]{h94}Hook, I.M., McMahon, R.G., Boyle, B., \& Irwin, M.J., 1994, \mnras, 268, 305
\bibitem[Ivezic et al. (2008)]{i08}Ivezic, Z., et al. 2008, arXiv:0805.2366
\bibitem[Jiang et al. (2006)]{j06}Jiang, L., et al. 2006, \aj, 131, 2788
\bibitem[Kaiser et al. (2002)]{k02}Kaiser, N., et al., 2002, Proc. SPIE, 4836, 154
\bibitem[Keller et al. (2007)]{k07}Keller, S.C., et al. 2007,PASA,24,1
\bibitem[Lawrence et al. (2007)]{l07}Lawrence, A., et al., 2007, MNRAS, 379, 1599
\bibitem[MacLeod  et al. (2011)]{m11}MacLeod, C.L.,  et al., 2011, \apj, 728, 26
\bibitem[Maddox et al. (2008)]{m08}Maddox, N., Hewett, P.C., Warren, S.J., Croom, S.M. 2008, MNRAS, 386, 1605
\bibitem[McDonald \& Eisenstein (2007)]{m07}McDonald, P. \& Eisenstein, D.J., 2007, Phys. ReV. D, 76, 063009
\bibitem[Palanque-Delabrouille et al. (2011)]{p11}Palanque-Delabrouille, P., et al., 2011, A\&A, in press (arXiv:1012.2391)
\bibitem[Peth, Ross \& Schneider (2011)]{prs11}Peth, M.A., Ross, N.P. \& Schneider D.P., 2011, \aj, 141, 105
\bibitem[Richards et al. (2002)]{r02}Richards, G.T., et al., 2002, AJ, 123, 2945
\bibitem[Richards et al. (2006)]{r06}Richards, G.T., et al., 2006, AJ, 131, 2766
\bibitem[Ross et al. (2011)]{r11}Ross, N., et al., 2011, \apj ~submitted, arXiv:1105.0606
\bibitem[Schlegel, Finkbeiner \& Davis (1998)]{s98}Schlegel, D.J., Finkbeiner, D.P., \& Davis, M. 1998, ApJ, 500. 525
\bibitem[Schneider et al. (2007)]{s07}Schneider, D.P., et al., 2007, AJ, 134, 102
\bibitem[Schneider et al. (2010)]{s10}Schneider, D.P., et al., 2010, AJ, 139, 2360
\bibitem[Schmidt et al. (2010)]{schmidt10}Schmidt, K.B., et al. 2010, \apj, 714,1194
\bibitem[Schmidt (1963)]{schmidt63}Schmidt, M., 1963, Nature, 197, 1040
\bibitem[Sesar et al. (2007)]{sb07}Sesar, B., et al., 2007, AJ, 134, 2236 
\bibitem[Sharp et al. (2002)]{s02}Sharp, R.G., et al., 2002, \mnras, 337, 1153
\bibitem[Skrutskie  et al. (2006)]{s06}Skrutskie, M.F.,  et al., 2006, AJ, 131, 1163
\bibitem[Smail et al. (2008)]{s08}Smail, I., et al., 2008, \mnras, 389, 407
\bibitem[Smith et al. (2005)]{sm08}Smith, J.R., et al.,  2005, MNRAS, 359, 57
\bibitem[Su et al. (1998)]{sc98}Su, D.Q., Cui, X., Wang, Y., Yao, Z.,  1998, Proc. SPIE, 3352, 76
\bibitem[The Dark Energy Survey Collaboration (2005)]{t05}The Dark Energy Survey Collaboration, 2005, astro-ph/0510346
\bibitem[Warren, Hewett \& Foltz (2000)]{w00}Warren, S.J., Hewett, P.C., \& Foltz, C.B., 2000, MNRAS, 312, 827
\bibitem[White (2003)]{w03}White, M., 2003, The Davis Meeting on Cosmic Inflation, p. 18, astro-ph/0305474 
\bibitem[Wolf et al. (2003)]{wc03}Wolf, C., et al., 2003, A\&A, 408, 499
\bibitem[Wu et al. (2010a)]{w10}Wu, X.-B., et al., 2010a, RAA, 10, 737
\bibitem[Wu et al. (2010b)]{wu10}Wu, X.-B., et al., 2010b, RAA, 10, 745
\bibitem[Wu \& Jia (2010)]{wj10}Wu, X.-B., \& Jia, Z., 2010, MNRAS, 406, 1583
\bibitem[York et al. (2000)]{y00}York, D.G., et al., 2000, AJ, 120,1579
\end{thebibliography}
\end{document}